\documentclass{ar-1col}
\usepackage[numbers]{natbib}
\usepackage{url}
\jname{Xxxx. Xxx. Xxx. Xxx.}
\jvol{AA}
\jyear{2019}
\doi{10.1146/((please add article doi))}

\begin{document}

\markboth{Unsleber and Reiher}{The Exploration of Chemical Reaction Networks}

\title{The Exploration of Chemical Reaction Networks}

\author{Jan P. Unsleber,$^{1,2}$ and Markus Reiher$^{1,3}$
\affil{$^1$Laboratory for Physical Chemistry, ETH Zurich, Vladimir-Prelog-Weg 2, 8093 Zurich, Switzerland}
\affil{$^2$ORCID: 0000-0003-3465-5788}
\affil{$^3$Corresponding author; e-mail: markus.reiher@phys.chem.ethz.ch; ORCID: 0000-0002-9508-1565}}

\begin{abstract}
Modern computational chemistry has reached a stage at which
massive exploration into chemical reaction space with unprecedented resolution with respect to the number of potentially
relevant molecular structures has become possible. Various algorithmic advances
have shown that such structural screenings must and can be automated and routinely carried out.
This will replace the standard approach of manually studying a selected and restricted number of molecular structures for   
a chemical mechanism. The complexity of the task has led to many different 
approaches. However, all of them address the same general target, namely to produce a complete atomistic picture of
the kinetics of a chemical process. It is the purpose of this overview to
categorize the problems that are to be targeted and to identify the principle components and challenges of automated exploration machines
so that the various existing approaches and future developments can be compared based on well-defined conceptual principles.
\end{abstract}

\begin{keywords}
chemical reaction networks, reaction space, automated network exploration, chemical processes, uncertainty quantification and propagation, kinetic modelling
\end{keywords}
\maketitle


\section{INTRODUCTION}
\label{sec:introduction}
Any chemical process may be decomposed in terms of a network of elementary steps.
The exact knowledge of all elementary steps, including intermediates, transition structures, 
and products allows for kinetic modelling and the prediction of concentration fluxes through the network.
Due to advances in the field of theoretical and computational chemistry in the past decades and the ever increasing computational power of modern hardware,
it has now become feasible to explore chemistry on a broad scale, i.e., tackling the vast dimension of
chemical reaction space. In general, the truly exhaustive exploration of some chemical process cannot be guaranteed, 
but the current state of theoretical and computational chemistry allows for the
generation of algorithms that march into an enormously larger fraction of this space than what would be accessible by manual exploration.

Chemical reaction space exploration comprises methods that generate knowledge about a chemical mechanism through atomistic modeling.
These methods can be closely tied to data-driven approaches, which attempt to induce reliable results about potential reactivity 
from existing experimental knowledge in the chemical literature.\cite{Corey1995}
Reactive pathways are then infered from rules and models generated on the basis of existing data\cite{Broadbelt1994,Broadbelt1996,Broadbelt2005}, such as
reactions published in patents\cite{Schwaller2018, Schwaller2018a} or curated databases\cite{Segler2018}.
These approaches were initially devised to exploit expert generated rules\cite{Pensak1977, Ihlenfeldt1996} and
have recently seen an impressive revival owing to the rise of machine learning 
techniques\cite{Gothard2012,Kowalik2012,Segler2017a,Coley2017a,Coley2017,Segler2017,Segler2018,Coley2018,Schwaller2018,Schwaller2018a}. 
Machine learning facilitates automating and abstracting the rule generation based on amounts of data that were unprocessable before.
This data-driven approach can be considered a powerful means to generate ideas about chemical reactivity for new systems similar
(according to some measure) to reactants studied in the literature. By contrast, chemical reaction mechanism exploration based on quantum-chemical
first-principles methods, which we consider in this work, provides an option, next to experimental synthesis, to probe the validity of
the ideas derived from data-driven inference, provided that depth, reliability, and accuracy of the quantum-chemical exploration
can be guaranteed. One aim of this overview is to work out the criteria that need to be considered and fulfilled in order to make
quantum-chemical mechanism exploration a reliable peer to data-driven reactivity deduction.

In a recent review\cite{Simm2019} we discussed in detail algorithms for chemical reaction space exploration. They may be grouped into three classes:\cite{Simm2019}
i) those that aim at a complete exploration of a given potential energy surface, ii) those that trade breadth for depth by relying on structure hopping 
and chemical heuristics, and finally, iii) those that exploit human intuition to tame the combinatorial explosion of structures in vast networks
involving numerous possible reactants and pathways through, e.g., steering by interactive quantum mechanics.
As further reviews on the topic can be found in Refs.~\citenum{Sameera2016,Dewyer2018}, we only mention some key methodological work 
in the field, highlighting that a variety of algorithms and concepts has already been devised.
They comprise graph-based approaches\cite{Habershon2016,Kim2018,Ismail2019}, first-principles heuristics that extract rules from the conceptual interpretation
of the electronic wave function\cite{Bergeler2015,Simm2017,Grimmel2019}, 
chemical heuristics using first-principles calculations\cite{Rappoport2014,Rappoport2019}, and stochastic approaches\cite{Kim2014,Habershon2015}.
It is not surprising that the global extension of local search and sampling methods has delivered an even broader range of exploration algorithms. Examples are
those that exploit artificial forces\cite{Ohno2004,Maeda2010,Maeda2013,Maeda2018}, growing string methods\cite{Zimmerman2015,Dewyer2017}, 
exploratory ab-initio molecular dynamics,\cite{Wang2014,Yang2017}
meta-dynamics\cite{Huber1994,Laio2002,Grimme2019,Rizzi2019}, 
and other enhanced sampling methods\cite{Martinez-Nunez2015,Varela2017,Yang2018,Debnath2019}. 

Since exhaustive overviews on the topic have already been provided\cite{Sameera2016,Dewyer2018,Simm2019},
this work will focus on comparability and missing links needed to make the different approaches comparable.
Because of the large number of concepts and algorithmic procedures introduced so far, it is rather difficult to have a balanced comparison of
the different exploration schemes on the same footing. This becomes even more difficult when considering that exploration algorithms
may be designed to serve specific purposes (e.g., gas-phase versus solution chemistry, restrictions to specific compound classes or computational
methods, and so forth).

There have been some attempts in the literature to compare existing algorithms at the example of specific target problems\cite{Grambow2018,Maeda2019}.
However, such comparisons are difficult. A trivial hurdle turns out to be the parameters and thresholds that control
an exploration algorithm, which may be chosen in a non-optimal way, making a direct comparison less conclusive.\cite{Maeda2019} 
Moreover, the performance of a some algorithm may also depend on the choice of a specific task. For this reason, it is desirable to have a set of 
criteria at hand, which allow one to arrive at informed and balanced conclusions about a specific computational approach.

For this reason, we adopt a meta-conceptual perspective aiming at the definition of common general concepts and requirements.
First, we discuss and categorize the range of possible mechanistic targets.
To keep this task well defined and controllable, we focus on the identification of elementary reaction steps and hardly touch upon the natural
extension and combination with subsequent kinetic modeling. 
For work relating to kinetic modeling in the context of reaction networks we would like to point the interested reader to 
Refs.~\citenum{Green1992,Susnow1997,Sumathi2002,Prinz2011,Sabbe2012,Bowman2013,VandeVijver2015,Gao2016,Wu2016,Proppe2016,Doepking2017,Han2018,Vereecken2018,Scherer2019,Andersen2019,Proppe2019,Cavallotti2019}.

\section{CATEGORIZING MECHANISTIC SEARCHES}
\label{sec:chemical_questions}
The central paradigm of reaction mechanism exploration is the idea that a chemical process under consideration
can be mapped onto a (transformation) network of elementary reaction steps connecting reactants and stable intermediates through
transition-state structures. 
As starting point, we may adopt the simplest concepts as they emerge from Eyring's absolute rate theory \cite{Eyring1935}
(or Kramers' theory \cite{Kramers1940, Haenggi1990}). For the sake of clarity, this may be considered sufficiently fundamental as a network of elementary
reaction steps can be arbitrarily refined through subsequent calculations that provide data required for more advanced rate theories.

\begin{figure}[h]
 \centering
 \includegraphics[width=.80\textwidth]{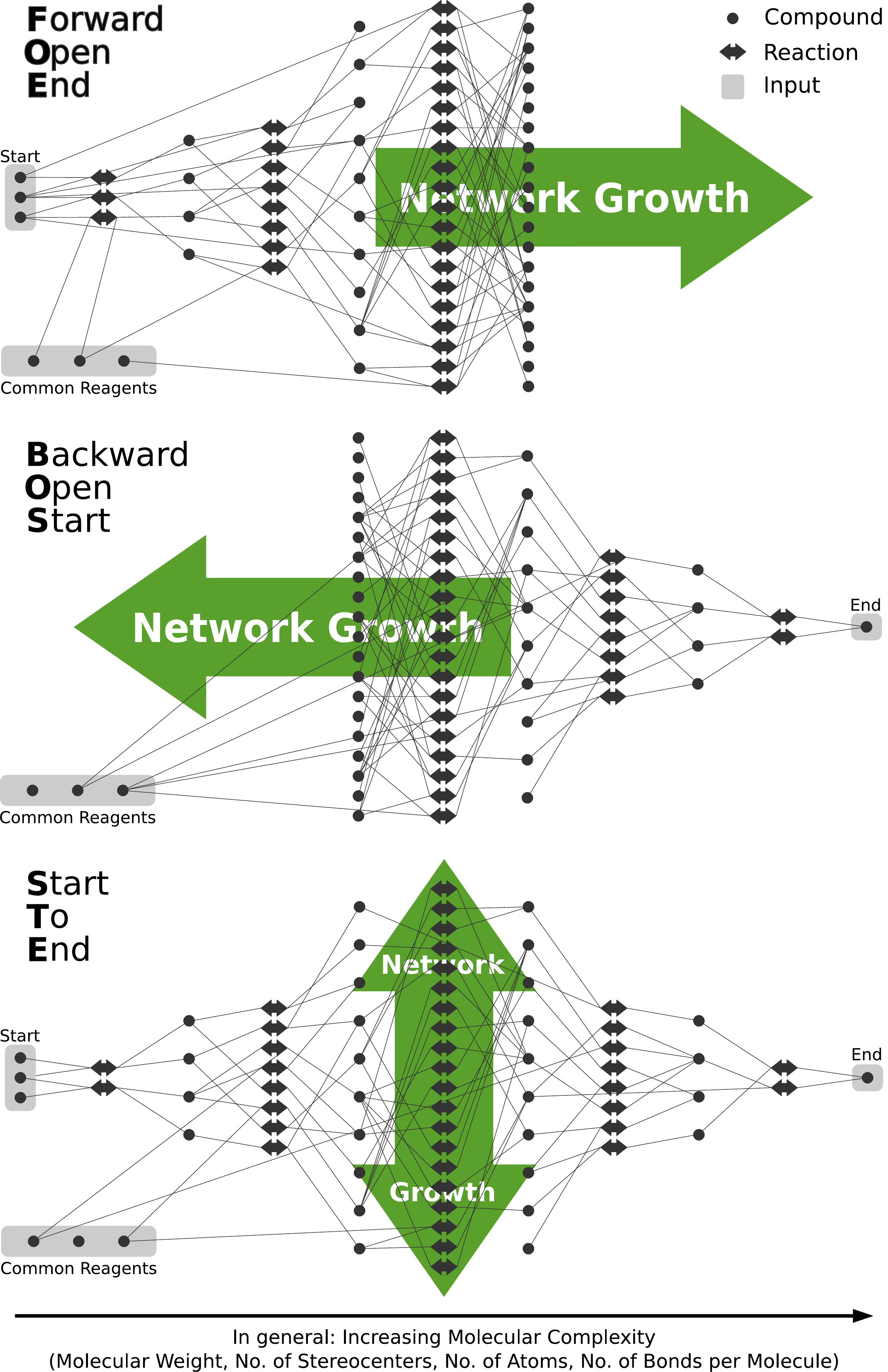}
 \caption{Three principal exploration types. 
Top: forward open-end (FOE) exploration; middle: backward open-start (BOS) exploration; bottom: start-to-end (STE) exploration.
Starting material (initial reactants) are given on the left-hand side and are typically of small molecular weight.
Common reagents (including ubiquitous molecules in the environment) are depicted separately to highlight their multi-purpose application.
A network can grow exponentially in forward explorations, because all newly emerging stable intermediates or side products need to be considered
as new reactants.
}
 \label{fig:network_growth}
\end{figure}

We consider all parts of chemical (reaction) space relevant for a problem as one huge network that encodes a chemical function.
Considering the fact that an exploration process will have some starting point and might already target at a specific end point, we
define three principle exploration types:
a forward exploration with an open end (FOE), a backwards exploration with an open start 
(BOS), and a start to end exploration with known start and end (STE) (see
{\bf Figure~\ref{fig:network_growth}}
for a graphical representation of these exploration types).

Asking 'How does compound A react with compound B?' manifests a prototypical forward exploration example with a potentially open end.
The compounds A and B constitute a known start. We must then assume that any stable intermediate of the detailed mechanism may, in principle,
again react with the starting compounds A and B, or with any other intermediate provided that it is sufficiently long-lived.
By contrast, the task 'How to synthesize compound C?' constitutes an inverse problem\cite{Weymuth2014,Sanchez-Lengeling2018,Freeze2019}, which will be difficult to solve
as the number and type of reagents in forward direction is not known in the beginning.
Hence, this backwards exploration of open-start type will in general require additional information on potential reagents and catalysts that
can promote the forward reaction. The machine-learning based retrosynthesis algorithms mentioned in the Introduction can provide
valuable hints on which reagents to choose.
If starting and end points of an exploration are for some reason known and fixed, an exploration algorithm may identify the viable
network of connecting elementary steps and then contribute to
the question of what is the (according to some measure) best pathway and how to promote it.

Due to the fact that the final product(s) of a chemical process may take part in subsequent reactions
and that chemical reactions are in general reversible, the labels 'start' and 'end' can become somewhat arbitrary.
Still, the definition of the three types discussed so far represent typical classes of exploration problems.

Although all reagents that are part of the network are treated in the same way, it can be advantageous to single out 
specific compounds due to their ubiquitous or universal nature as reactants such as, for example,
H$_2$O, $^1$O$_2$, small elimination products (e.g., HBr), (solvated) protons and so on.
In this way, screening tasks such as those for a reaction's robustness towards functionalized
molecules\cite{Collins2014} can be simplified. Further conceptual steps, such as work-up, reactant separation, and sequential reaction steps can be easily
encoded into a network by restricting the connectivity of certain nodes.
Note, however, that a specific reference to an actual environment of a reaction is not made in the labels introduced so far.
Implicit and explicit solvation, as well as metal surfaces, protein environments, or other environments will initially lead to different, though related networks.

The setting defined so far comprises a broad range of scenarios from 
simple reaction mechanisms in the gas phase (STE without additional common reagents) to auto-catalyzed reactions with side reactions
(FOE with additional common reagents), to probing a restrosynthetic proposal for synthesis (STE with various restrictions and constraints) 
to screening of an optimum catalyst for a specific target transformation (STE with different catalysts as common reagents).
Classifying  reaction network exploration protocols according to the three types introduced above allows us to categorize the different algorithmic
and conceptual approaches towards the exploration challenge and to compare them on the same basis.
Considering the open nature of  FOE-type explorations,
it is obvious that an actual algorithm will require some adjustable termination determination of the exploration.
For the termination of FOE and BOS explorations, bounds on molecular weights are simple criteria and in the case of STE explorations the number
of reactions between start and end could be limited.
Furthermore, constraints based on kinetic modelling at a certain temperature and other external constraints can be introduced.

\section{GENERAL NOMENCLATURE FOR REACTION NETWORKS}
\label{sec:nomenclature}
Since notions such as  'system', 'molecule', 'compound', 'reagent' and 'structure' are often used interchangeably, we are advised to 
introduce and define the essential concepts.
To define the components of a reaction network, we first settle on the notion of a 'molecular structure':
\begin{description}
 \item [Molecular structure] 
   A single arrangement of atomic nuclei on a Born--Oppenheimer potential energy surface shall be denoted a molecular structure.
   As a consequence, it is represented by a set of Cartesian coordinates for the nuclei. Depending on the interpretion in terms of
   a kinetic theory, we will
   y, in general, be interested in structures with specific properties, most importantly those with vanishing
   geometry gradients that highlight stable intermediates and first-order transition states. 
 \item [Properties] 
   Each such molecular structure may then be assigned
   some property that relates to an electronic structure defined by the number of electrons and spin assigned, its state of excitation (typically
   the ground state) and any derived molecular property (such as the electronic energy or total dipole moment).
   Here, we may also allow for conceptual properties such as partial charges and local spin that rely on some chosen
   decomposition scheme.
\end{description}
With these basic terms we may proceed and introduce a 'compound' as
\begin{description}
 \item [Compound]
   A set of molecular structures with the same nuclear composition and connectivity in terms of chemical bonds
   denotes a chemical compound. The bonding pattern that is crucial for the assignment to the set may either be
   fixed based on heuristic rules or, preferably, be determined through a bonding analysis (e.g, in terms of Mayer
   bond orders\cite{Mayer1983}). 
\end{description}
While such a bonding analysis may often be unique (and particularly helpful to detect and assign molecular strutures as
different conformers to the same compound), it cannot be excluded that certain structures end up in a gray zone (for instance,
if bond orders are very low). This may then create ambiguities on the conceptual level of 'compounds', but not on the more fundamental 
level of the molecular 'structures' (characterized by a vanishing gradient).

Whereas we consider 'structures' essentially as static for quantum-chemical  single-point energy calculations, the concept
can be extended to molecular dynamics simulations. 
Then, however, clustering algorithms\cite{Karpen1993,Shenkin1994,Jain1999,Shao2007,Keller2010} will be required to assign sets of structures
to a node, which is then a generalization of the 'compound' concept. Transitions between such compounds derived from clustering trajectory
data can provide access to kinetic information through Markov state modelling\cite{Singhal2004,Bowman2009,Prinz2011,Li2016}.
A single-point-calculation approach will require the explicit generation of all relevant conformers (for example, with the program
package RDKit\cite{Landrum2019}, whereas molecular dynamics simulations then generate compounds through clustering.

\begin{figure}[h!]
 \centering
 \includegraphics[width=1.00\textwidth]{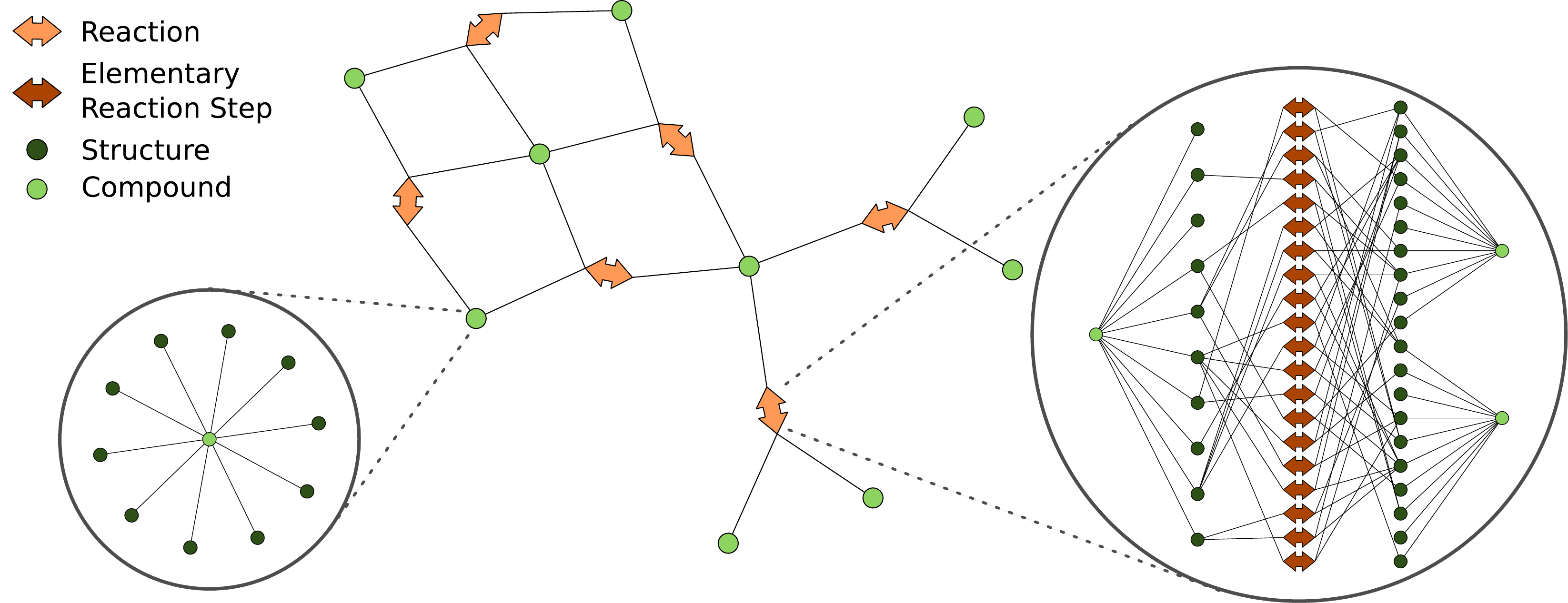}
 \caption{Example network built from reactions and compounds, expanded section shows a sub-network structures which highlight 
          the relation between structures and compounds as well as the relation between reactions and elementary steps.}
 \label{fig:network}
\end{figure}

In order to generate relations between structures and eventually compounds, transformations from one node in the network to another one
are to be uncovered. In general, one will first strive to identify all relevant elementary steps that connect molecular structures
and then later generate a reaction network out of these elementary steps that transform compounds into one another. 
For transformations we will use the following terms:
\begin{description}
 \item [Elementary (Reaction) Step]
   One (or more) structures may be connected with one (or more) different structures via a single 
   transition state representing an elementary step that can be resolved as a sequence of structures
   along a corresponding minimum energy pathway.  
 \item [Reaction]
   A chemical transformation of one (or more) compounds into one (or more) different compounds represents a reaction that
   can be distilled from at least one, but in general more than one elementary steps. 
\end{description}
As an example, {\bf Figure~\ref{fig:network}} shows a small reaction network built from the four types of nodes defined above.

Any algorithm tackling a chemical question posed in terms of changing molecular structures
should explore the transformation steps of these structures  and eventually generate compounds and reactions. Hence, chemical
reactivity and function will then be encoded in terms of the elementary steps connecting structures as nodes of a network.
However, we may introduce a few more general concepts that allow for comparison between compounds and reactions. These 
generalizations are important to cover concepts such as 'residue' that assign a spectator role to certain parts of a molecule,
while they are indispensable in an explicit quantum chemical description, which, in general, requires detailed knowledge about all electrons and nuclei
involved. As such, these generalizations represent the first step towards mechanistic interpretation, because the network of structures and
their connecting elementary steps is agnostic with respect to the assignment of any role in the chemical function under consideration. 
Moreover, generalizations can be invoked to assign networks that are very different on the level of atoms and elementary particles 
to the same class. Typical examples are generalizations of molecular function to be fulfilled by a comparatively wide range
of compounds, such as are 'oxidizing agent' or 'hydride source'.
For these reasons, we introduce three additional terms:

\begin{description}
 \item[State of a Compound]
   Reducing or oxidizing a compound (and hence, all of its structures), changing the degree of excitation or the spin state, or 
   protonation lead to different molecular structures in a quantum chemical description and would therefore produce different nodes
   in a reaction network from which new elementary steps lead to further structures. The close relationship to quantum mechanical concepts
   and prototypical reactions (protonation, electron transfer) make them amenable to automated identification and assignment in quantum
   chemical procedures.
 \item[Reagent]
   A group of compounds that shares the same role in a network form the category of a 'reagent' that is specific for that role. 
   Typically reagent are 'oxidant', 'reductant', 'acid', 'base', and 'catalyst'. 
 \item[Purpose]
   The concept of 'reagent' allows for comparison and a higher degree of abstraction of a network, which for
   the network itself is its 'purpose'. 'Purpose' is the superordinate concept to 'reaction'.
   A typical 'purpose' in reaction chemistry is 'synthesis' or 'catalysis', but the term may as well refer to any molecular
   functions encoded in terms of structural changes such as 'molecular motor' or 'mechanochemical device'.
\end{description}

To assign an oxidation, charge, spin, excitation, protonation state to a compound not only attributes chemical meaning to a network
of structures, it also has consequences for network data handling, storage, analysis, and presentation. Given the very fundamental nature
of these states also allows for their monitoring during exploration.\cite{Vaucher2016a}
Moreover, note that some compounds may be part of different sets of 'reagent' in different contexts.
Early stage exploration algorithms will not feature these last two types of nodes and how they encode 'purpose', because of the bottom-up
nature of reaction network exploration based on the first-principles of quantum mechanics.

Given two or more explored networks, these concepts define connections between them, which can and should be exploited
by exploration algorithms. One type of connection is node sharing,
i.e., the fact that a compound (and hence, its structures) can occur also in another network. Calculated data on such
parts of a network can be re-used and call for a centralized database storage (the \textsc{QCArchive} project by \textsc{MolSSI}\cite{qcarchive} is one example
for a technical realization that could be exploited by network exploration algorithms). Another type of connection is contextually joining 
reaction networks of structures that are very different on a quantum chemical level 
through the assignment of the same purpose. A typical example is a generic catalytic cycle, which on a realization
level can put networks with different catalysts or with differently substituted substrates into the same context. 
As a result, one could even derive functional networks in which modulating effects of molecular structure are hidden in an environment
description of a quantum-chemical embedding approach.

Deliberately, we have left some terms unspecified in order to allow for some vagueness in the context of network exploration.
This is convenient as it facilitates comparability of different algorithms and does not interfere with the core concepts introduced above.
For instance, we have not specified the term 'system' and may use it as a stand-in for the way possible elementary steps are actually set up and then
explored in an algorithm.

\section{FEATURES OF REACTION NETWORK EXPLORATION ALGORITHMS}

\subsection{General Foci}
The main results of a reaction network exploration will be extracted at the level of compounds and reactions.
Context-based abstractions into reagents and purpose then allow for arranging and highlighting the vast amounts of data.
Splitting up reactions and compounds into structures and elementary steps allows for understanding the fine details of a particular mechanisms 
at the level of the physical dynamics of the involved molecular species.

At the level of compounds and reactions, it is reasonable to distinguish between two dimensions:
the breadth of a network and its depth. 
Here, breadth refers to the amount of reactions and compounds that have been incorporated into the network, whereas
depth denotes the amount of structures and elementary steps discovered for each of the compounds and reactions.
Naturally, both will be difficult to determine in absolute terms as this would require complete knowledge about a
chemical process, which is actually the target of an exploration algorithm.

A lack of depth will likely yield qualitatively wrong kinetics, and hence, it will predict wrong distributions of reactants across the networks
once these are modeled.
Depth fidelity also comprises the accuracy of the chosen computational methods which manifests itself in errors on barrier heights.
A lack of network breadth will yield qualitatively wrong results for the total kinetics as it would imply that important side reactions have been missed.

In order to describe a chemical process as reliably as possible, both dimensions have to be pushed to their limits and it will be important
to find algorithm-intrinsic measures that hint toward their saturation.
Therefore, it is instructive to define two distinct types of completeness, one for the breadth, the graph fidelity, and one for the depth, the node fidelity.

Naturally, different applications can require a different focus with respect to these directions.
Consider the following two examples:
\begin{itemize}
 \item The calculation of the feasibility of a single reaction cascade to gain insights 
       into its reaction mechanism by exploring a multitude of conformers and pathways
       requires slow growth of the network breadth and very accurate exploration of its depth.
 \item The exploration of an entirely unknown set of reactions of given compounds may favor a 
       quick growth, with an a posteriori automatic refinement scheme to increase depth and breadth for 'interesting' parts of the network.
       A low accuracy in terms of depth for unaccesible and therefore uninteresting parts of the network
       will be sufficient and often mandatory for the sake of feasibility.
\end{itemize}

As should have become apparent by now, there exists a multitude of requirements for a generally applicable exploration tool.
It shall be able to switch between the different exploration modes defined in Section~\ref{sec:chemical_questions}
and also arrive at a sufficient accuracy.
To this end, we define a more complete set of key challenges for exploration algorithms in the next section that
allows one to classify and assess algorithmic developments regarding their scope and capabilities.

\subsection{Challenges}
\label{sec:goals}
In order to later compare and eventually rank algorithms that explore chemical space, computationally it 
is imperative to define their goals. 
These targets presented in the next subsections are driven by the following general challenges:
\begin{description}
 \item [Validation Challenge] 
       The exquisite details that exploration algorithms can generate for any chemical process raise the question of reliability
       as, in general, no or very little experimental or theoretical reference data will be available. As a consequence, uncertainty
       quantification\cite{Simm2017a} will become a crucial part of the whole exploration process\cite{Proppe2019,Simm2018}.
 \item [Operating with Huge Amounts of Raw Data]
       Automated exploration algorithms will be most useful for cases that require thousands or millions of structure searches and optimizations.
       As a consequence, huge amounts of data will be produced whose manual inspection is not at all possible. As a consequence, automated
       exploration algorithms must be very stable, handle all data in a fully automated and integrated way, and 
       automatically draw the operator's attention to critical situations such as convergence failures that cannot be resolved automatically.
 \item [Minimal Expectations on the Operator Side]
       The software design that implements an exploration algorithm must take into account that all its parameters, thresholds, screws, and bolts
       cannot be fully understood on the application side. In other words, the efficiency and reliability of the algorithm should depend as little
       as possible on knowledge about the intricate effects of changing some of its parameters. Naturally, the default values for them should be
       expected to be stable and applicable to a wide range of problems and situations.
 \item [Unknown Degree of Incompleteness of Generated Data]
       For any reasonably complicated case, it cannot be rigorously proven that the exploration of the corresponding reaction network of interest
       has been completely explored. As a consequence, it will be difficult to construct sufficiently complex benchmark cases against which exploration depth and breadth can
       be measured, and most likely, these will only emerge in a joint effort of various approaches over time.
\end{description}

\subsection{Targets}
Within this basic setting, we may now formulate goals and targets of exploration algorithms.
Some of these goals have been tackled already, but all of them remain to be improved upon in some ways. 
It is therefore decisive to have this list at hand for future developments in the field.

\subsubsection{General Applicability and Stability}
A key goal for the exploration of reaction space must be flexibility with respect to the class of molecules and their environments that can be considered.
An algorithmic restriction towards a specific compound class, reaction environment, or state of aggregation can be a severe limitation in the
discovery process as it would severely limit the domain of exploration and exclude potentially decisive reagents, solvents, and so on. 
An ultimately useful protocol must be able to accommodate any potentially relevant molecular scenario, even if it is truly hard to
map in a virtual screening process. Examples can be found in structures with  multiple transition metals in a protein environment or with
molecules on metal surfaces. 

Moreover, it must be ensured that the algorithms employed are as stable as possible because an exploration protocol may lead them to regions
of configuration space that pose problems for them. For instance, orbital convergence will be a key issue -- solution methods
for complete failure of convergence\cite{Wang2019} or for convergence to wrong solutions\cite{Vaucher2017} are registered.

It will, in general, be necessary to operate with a wide range of approaches, spanning fast and less accurate as well as slow and accurate 
energy assignment protocols, from detailed structure construction and search to advanced sampling through molecular dynamics and Monte Carlo methods.
Therefore, the requirements for exploration algorithms in terms of general applicability are immense.

\subsubsection{Intrinsic Constraint Monitoring and Adaptation of the Exploration Algorithm}
Obviously, any algorithm will involve choices that eventually limit its applicability. For instance, exploring reaction space based on individual evaluations of the
stationary Schrodinger equation emphasizes the role of the electronic energy, which will be reasonable as long as it determines the major part
of the relevant energy differences. When the exploration then enters a regime of a potential energy surface that is rugged, then this needs to
be substituted by a proper molecular dynamics or Monte Carlo sampling approach -- and vice versa.
The quantum chemically explored reaction network can then be supplemented with proper information from some enhanced sampling approach,
possibly encoded in terms of a kinetic model built from this sampling (e.g., through Markov state models\cite{Singhal2004,Bowman2009,Prinz2011,Li2016}).

In order to recognize these limitations, it is crucial to determine the limits with algorithm-intrinsic means
while exploring a network in order to achieve sufficient breadth and depth. In the example above, a large number of small barrier heights 
in some region of reaction space can be taken as a sufficient indication. 
As long as no fully satisfactory integrated software is available, interoperability of different 
implementations will be of key importance.

\subsubsection{Taming Conformational Explosion}
With increasing size of molecular structures, conformations of a compound become increasingly important. For moderately sized
compounds, conformers can be explicitly constructed\cite{Hawkins2017,Simm2017,Landrum2019} and optimized. At some point, however, this will no longer be feasible, and again,
sampling approaches will be needed. For truly large molecules, this will eventually become a cumbersome task as highlighted by the
protein folding problem.
The whole menace of conformational depth becomes even more severe when considering transition state structures, especially when embedded 
into a fluxional environment such as, for instance, water.
An exploration based on a single conformation per compound could be a first step to generate an overview on the most interesting parts
of a network of elementary steps, but later refinement will then be more mandatory.

\subsubsection{Type of Energy Data Provided}
It is obvious that ultimate energetical data assigned to nodes of a network should be free-energy data within a well defined thermodynamic ensemble.
However, this ultimate goal is not easily accessible to arbitrary accuracy and a first step would therefore be a network based on electronic energies only
(to be later refined by modelling contributions from the nuclear framework at some finite temperature). Given the fact that macroscopic constraints such
as temperature, pressure, volume, particle number, and so forth can change, it would be desirable to store all data per structure from calculations, which allow
one to evaluate free-energy data for changing external parameters at will.
Arriving at free energies may be accomplished within very different models ranging from those that start with the standard
rigid-rotor harmonic-oscillator particle-in-a-box model (see, e.g., Ref.~\citenum{Proppe2016}),
to continuum solvation approaches (see, e.g., Ref.~\citenum{Klamt1995}),
to additivity schemes (see, e.g., Ref.~\citenum{Grambow2019}),
and to explicit sampling approaches (see, e.g., Ref.~\citenum{Sidler2016,Sidler2017}).

\subsubsection{Environment Embedding}
Exploring some chemical function or reactivity is most easily accomplished for reactants lacking any environment (i.e., for isolated species that may be considered
to represent a gas-phase situation at low pressure). The inclusion of a suitable environment represents a major challenge that
must be addressed by an exploration algorithm. Clearly, explicit molecular dynamics in a box of sufficient size under periodic boundary
conditions is an attractive choice, but quickly limits the scope of an exploration to the Born-Oppenheimer surface that can be constructed for
the elementary particles in that specific box. Changing reagents will not easily be possible. Hence, embedding schemes (for reviews see Refs.~\citenum{SeveroPereiraGomes2012,Jacob2014,Wesolowski2015,Lee2019})
play an important role, possible combined in a multi-layer strategy that extends from explicit environment structures
close to a reactant to structureless dielectric environments at large distance.
Different embedding scenarios will lead to variations of similar networks that require specific care on the data-management size in order to allow
for relevant chemical interpretation and conclusion.

\subsubsection{Error and Uncertainty Diagnostics}
Any model that produces raw data for a network exploration algorithm will rely on certain approximations. For instance, electronic structure models
are likely to be based on fast semiempirical\cite{Husch2018} or density-functional theory\cite{Parr1994} methods, which can be affected by surprisingly large
errors for specific structures (see  Refs.~\cite{Weymuth2014a,Simm2016,Husch2018a} for examples). Without proper uncertainty quantification, explorations
based on such data will be inherently unreliable. Error estimation\cite{Simm2017a} therefore becomes key.
Methods have been devised that point the way of how this can be achieved.\cite{Pernot2015,Proppe2016,Pernot2017,Proppe2017,Simm2018}

\subsubsection{Automated Error Reduction}
Naturally, nonnegligible errors for certain structures detected during exploration require (automated) refinement by launching more accurate energy calculations. 
Typically, this will encompass starting reliable ab initio calculations. While this is feasible for single-referece methods such as explicitly correlated
coupled cluster theories \cite{Ma2018}, which can typically be run as black boxes, this is not that straightforward for multi-configurational
problems although significant progress has been made in this respect \cite{Stein2016a,Stein2016,Stein2019}. Then, only feasibility consideration
are an issue, but may be alleviated by embedding calculations\cite{Manby2012,Tamukong2014,Hegely2016,Muehlbach2018,Lee2019}.
If an exploration protocol allows for such calculations on demand, then the determined error should be propagated through the network in order
to reduce the uncertainties for the approximate exploration method.
As we have shown\cite{Simm2018,Proppe2019a}, this is possible by means of machine learning.
Such approaches are determined to play an important role in maximizing the accuracy whilst minimizing the computational cost of the network explorations.

\subsubsection{Intuitive and Immersive Interaction and Visualization}
As vast networks will contain too much data for a human to grasp, it is imperative that results of any exploration can be
displayed in an accessible way and that conclusions can be drawn with the help of algorithms. This requires specific software for human-machine
interaction as visual inspection of alphanumerical data will be unfeasible and pointless.
Suitable graphical user interfaces are required that also free one from looking at alphanumerical raw data stored in some databank container.
They may invoke new hardware (see, for instance, Refs.~\citenum{Haag2014,Haag2014a,OConnor2018,Amabilino2019}) to intuitively
experience these data in order to easily put focus on relevant aspects. Such hard- and software also allows for manual interference and control of
the otherwise automated exploration process.

Obviously, interaction cannot mean manual editing of input files for a specific program and a manual database query and insertion, but rather
a simple click on a node, opening a context-dependent minimal menu.
In view of the definitions introduced above, it is possible to define layers for this purpose reflecting in the first layer
structures and elementary steps, in the second compounds and reactions, and then on top of this one abstract layer that encodes reagents and purposes.
{\bf Figure~\ref{fig:network}} shows one example of such a representation, also indicating the idea of switching between these layers.

It appears natural that interacting with such an interface and steering the whole exploration protocol would be natural (and for nonexperts most
convenient) through language processing algorithms. 
Voice control of exploration is one option that has already been proposed as a valuable goal for computational chemistry.\cite{Lu2017,Aspuru-Guzik2018}
Given the advances in the context of daily-life artificial-intelligence assistants and the availability of open-source language processing libraries \cite{Fast2017,mycroft,jarvis}, 
this tasks is not at all difficult and even the set-up of an interpretation assistant, which can translate between chemical jargon and the precise ingredients
required for quantum chemical calculations on a properly prepared atomistic model, will be rather straightforward.

\subsubsection{Maximum Accessibility}
While many quantum chemistry programs are reasonably easy to install, an exploration software will most likely involve a database, a user front-end, 
and a back-end that manages all calculations.
Hence, installation and therefore accessibility are more complex by design.
Both hardware and quantum chemical software packages available at the outset of an exploration attempt will result in various setups
with different computational capabilities.
Given the recent rise of cloud computing and virtual machines an easy access to standardized configurations of the back-end accessible as images 
is likely a key feature that allows easy setup of the entire machinery for nonexperts.
While proprietary software can be made accessibe in this way, open-source initiatives are likely to form a more stable and
user supported base. The latter are to be prefered in any case due to better reproducibility, reliability, and fidelity.
With the plethora of settings already available in for quantum chemical calculations and the profound impact they have 
on the resulting data\cite{Lejaeghere2016} it will only be possible to reproduce and understand data if their production and processing can be 
inspected at the source-code level.
Given the additional algorithmic layers needed in a reaction network exploration it is of key importance to document and explain
possible settings. 
For the same reasons a comprehensive way of accessing and documenting the used options for a given exploration run is another key part of this goal.

\subsubsection{Data Transferability}
For efficient explorations it is highly desirable that already generated networks and the contained data can be reused and combined such that 
subsequent explorations do not need to recalculate them and instead can incoporate existing data seamlessly. 
Considering the vast amount of standard reactions and the resulting list of frequently used reagents and often occuring reactants, data integration 
will be highly efficient.  However, integration of existing data
will only leverage an exploration attempt if uncertainty quantification has rated and labeled the existing data.
It is therefore desirable to generate a central library or database of highly accurate results that is continuously extended with data from local explorations, if
that data has been generated with an accuracy above a certain threshold.
A central database of chemical reaction space would also be a promising starting point for the application  of meta-algorithms and machine
learning models to exploit and learn from the reaction chemistry mapped out.

\subsubsection{Enhanced Kinetic Modelling}
Eventually, kinetic modelling will be required to study concentration fluxes through a network. As it is a priori not clear what the number and kind of elementary steps
will be that constitute the network, rather general microkinetic solvers will be required, possibly tightly entangled with the exploration algorithm itself
to enhance and guide the latter.\cite{Proppe2019}
As a consequence, such solvers should be capable of dealing with vast time-scale ranges and varying degrees of molecularity (at least up to second order kinetics).
Whereas special tools\cite{Glowacki2012} and commercial solutions\cite{Kee1980} are already available, it is obvious that reaction network exploration
presents further challenges that demand more developments.
Moreover, to accurately model the flux in reaction networks, theories beyond Eyring's absolute rate theory 
will be needed.\cite{Miller1993,Truhlar1996,Pollak2005,Garrett2005,FD2016} 
Clearly, also quantum tunneling must be considered\cite{Richardson2018}. However, improvements on this level will represent a natural extension
of a network of elementary steps as the software may automatically gather more information about the potential energy surface in the vicinity of the
nodes in such a way that advanced rate theoretical expressions can be evaluated.

\section{OPTIONS FOR COMPARISONS OF NETWORK EXPLORATION ALGORITHMS}
\label{sec:comparisons}
Comparisons of existing algorithms that explore reaction networks have already been attempted.\cite{Grambow2018,Maeda2019}
However, in view of the many aspects to be considered in such an attempt and the fact that a balanced set of benchmark reaction networks
resembling various scenarios of practical relevance would be required, demonstrate that such comparisons will not be easy, although certainly
needed.  It is important to assemble a set of criteria by which an exploration algorithm should be assessed and rated in order to
highlight the different problem classes that need to be addressed. It will be necessary to create a multi-dimensional
diagnostic to characterize a specific exploration protocol.

Ultimately, any such measure will have to be based on solid comparable data, and hence, benchmark data should be generated and collected in future work.
Given the many different types of chemical processes that could be probed, an extendable set of smaller networks appears to be the best choice in practice.
Using the three initially defined exploration modes (FOE, BOS, and STE; see Section~\ref{sec:chemical_questions}) we may propose three main tests 
to be run for each benchmark network.
In view of the fact that automated network exploration has a significant and important technical and implementational component,
extensibility and access should be measured.
Furthermore, the reproducibility in terms of the licensing of the source code should also be a factor.
Assuming a scale from one to ten, a software that is closed source and requires a specific set of commercial software without any
possible extension would be ranked zero.
A fully open-source software with the possibility to interface any other quantum chemical code and a set of open-source
default programs provided with the exploration tool to guarantee full functionality would rank as ten.

\begin{figure}[h]
 \centering
 \includegraphics[width=1.00\textwidth]{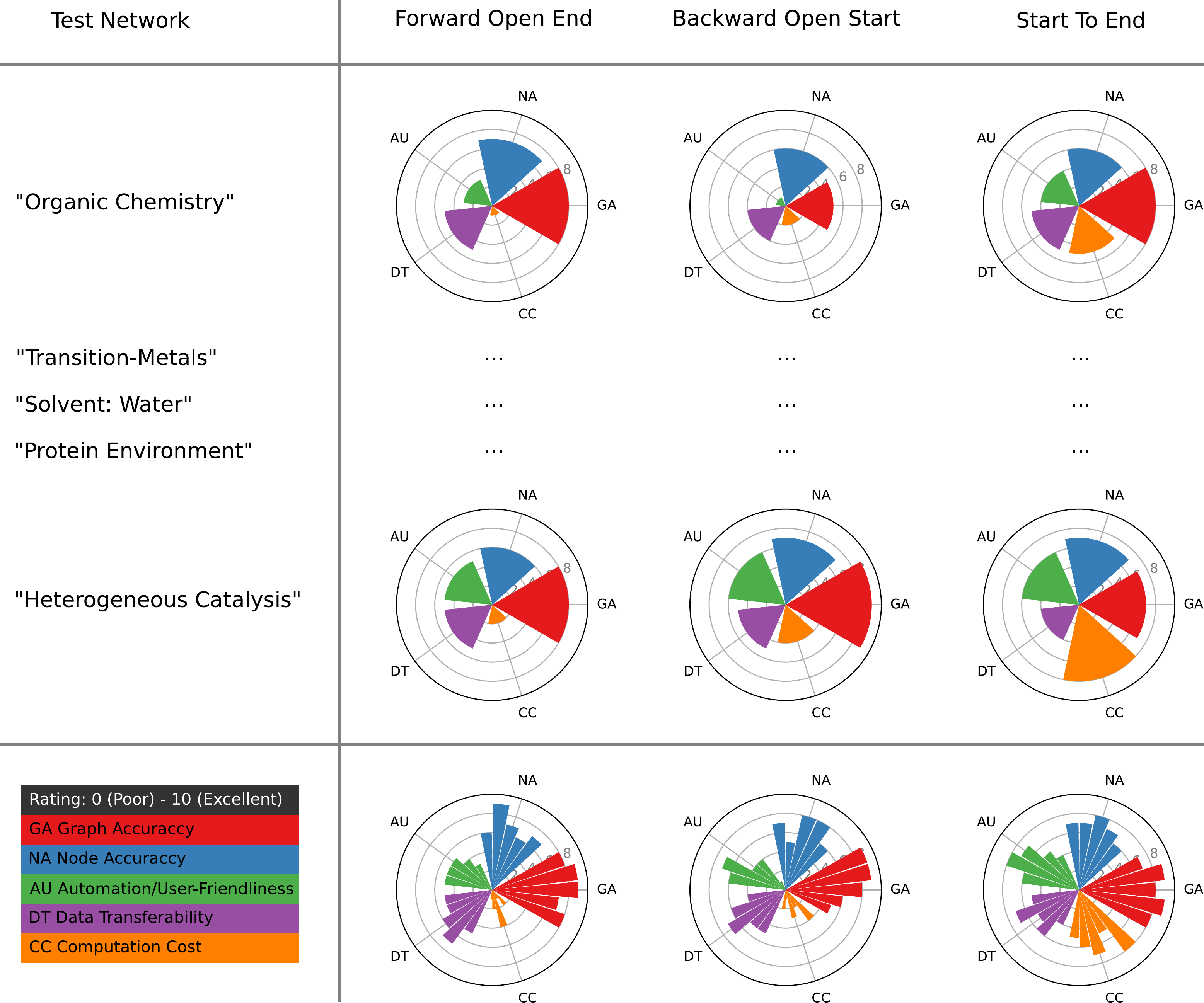}
 \caption{Fictitious benchmark data shown in rose plots for a chemical reaction network exploration software.}
 \label{fig:example_radar}
\end{figure}

For a fine-grained comparison of exploration algorithms and implementations assessed at benchmark networks five key descriptors sketched in
{\bf Figure~\ref{fig:example_radar}} should be sufficient.
The five descriptors first and foremost include the two types of reliability measures described above.
The node fidelity (depth) is calculated by comparison of the energy spectra of the different nodes with the reference data.
The network fidelity (breadth) is calculated as the amount of correctly identified compounds and reactions.
Naturally, the expense of the calculations is also one of the descriptors.
The computational cost could be measured similar to the work in Refs.~\cite{Grambow2018,Maeda2019}, which uses the number of gradient 
calculations as dominant measure.
Due to the fact that the accuracy depends on the methods used and that gradient calculations with different methods 
can vary significantly in duration, the measure should also incorporate an average time of a gradient calculation
on a certain hardware.
In order to probe the transferability of networks, meaning the possibility to reuse previously generated network data,
we may measure the acceleration of an exploration when starting from a subset of already existing nodes in the reference network
and also supplementing the running exploration on the fly with structures surfacing during the exploration that are found to be already
in a central database.
Furthermore, nodes which are automatically determined to be similar (e.g., nodes that were calculated with different solvents as environment)
can and should be used as starting guesses.
In order to credit the fact that these software packages and algorithms should eventually be useful to non experts in the
field of computational and theoretical chemistry, it is important to measure the user-friendliness (immersion ) and also the degree of full automation
of a given protocol.
The amount of human work required to steer and control an automated network exploration software could be taken as a measure for user-friendliness and for the level of 
automation achieved. Unfortunately, this may require extended users studies.
Given the vastly different approaches to actually solve many of the problems discussed in this work, a challenge similar to the CCDC organic crystal structure prediction challenge\cite{Reilly2016} 
or the SAMPL challenge\cite{Rizzi2018} could be a desirable testing ground for the field.
It would allow for better comparisons for subtopics and also help focus the efforts on particular unsolved problems.

\section{ASSESSMENT OF THE CURRENT STATUS}
\label{sec:current_status}
So far, we have discussed general concepts and goals for reaction network exploration algorithms and their implementation.
To assess the current status of the field we will highlight some aspects according to the literature.
First, it is noteworthy that most exploration methods employ existing quantum chemical software packages and models.
They generally do not implement new electronic structure models in a closed-source form, which has advantages for availability, reliability, and reproducibility
of the overall method.
The rigorous estimation of the associated errors on structures and energies, their propagation and minimization during the exploration 
have been tackled only in a few instances.\cite{Proppe2016,Simm2017a,Doepking2017,Simm2018}
In terms of the scope of the chemistry that has been the subject of explorations attempts so far, most methods work more or less routinely for basic (gas-phase) organic chemistry, and
some of the reported algorithms have been shown to work with transition metals\cite{Maeda2012,Varela2017,Bergeler2015,Habershon2015,Sameera2016,Habershon2016,Kim2018,Grimme2019,Grimmel2019}.
The explorations reported so far have largely avoided explicit description of environments such as solvents or protein embedding.
Most applications reported for exploration algorithms have been of the proof-of-principle type. 
It is therefore still a long way ahead of us until we have general implementations available that can be routinely employed in daily research.
No study has reported data on a network with, say, $1,000$ confirmed compounds and $10,000$ possible reactions. 
(Note that this implies $\gg1,000$ conformers to be generated and $\gg10,000$ elementary steps to be calculated.)
The challenges of the visualization and analysis of such an amount of data and the transferability of sub-spaces of the network 
have only been touched upon\cite{Bergeler2015,Simm2017}.
Overall, however, important developments have been accomplished and key steps haven been taken (see Sec.~\ref{sec:introduction}).
Considering three major requirements, i.e., a robust yet flexible exploration algorithm, a scalable and extendable back-end for the actual computations,
and kinetic modelling relating the generated data to actual chemistry, we note that no single software framework has been established so far
that accomplishes these goals so that comparisons of implementations on equal footing become feasible.
We note, however, that we have set out to provide such a program package\cite{scine}.

\section{CONCLUSIONS}
\label{sec:conclusion}
In this work, we provided a general description for automated atomistic reaction network exploration algorithms. Owing to the huge diversity
and heterogeneity of the tasks, we considered it necessary to define core concepts, targets, and challenges in order to make this fast growing
field accessible to assessment and validation and to identify weak spots in existing schemes. Although a detailed rating of existing schemes has
been beyond the scope of this work for various reasons, of which the limited accessibility of implementations is an important one, the present overview
may still be very well suited as a general guideline toward the application range and capabilities of existing algorithms.

We defined three basic exploration patterns for chemical reaction networks that depend on the scientific purpose for which they are designed.
In connection with a basic nomenclature presented afterwards 
we then introduced two fidelity measures related to the breadth and depth of a network. 
These are  graph accuracy, i.e., the correct identification and connectivity of nodes, and node accuracy, i.e., the correct description of the 
set of structures that represent a compound. We continued to classify prototypical tasks for network exploration algorithms.
These include a high level of automation, error diagnostics and automated uncertainty handling, 
network data transferability, and eventually kinetic modelling. 
Finally we proposed a transferable comparartive scheme of reaction space exploration software based on the concepts and targets elaborated on
in this work.
Key measures will be: node and graph accuracy, software extensibility, automation, immersion, and user-friendliness, data transferability, and of course, 
the computational cost associated with a given algorithm/software.

We are certain that such a multi-dimensional ranking wil become decisive in future work on reaction network exploration algorithms
as it can advance the field by clear categorization of new exploration protocols in the context of existing ones. Detailed quantitative
measures can then be provided for existing and new schemes once balanced benchmark networks become available that represent the
variety of algorithmic features highlighted in this work.



\begin{summary}[SUMMARY POINTS]
\begin{enumerate}
\item Predictive theoretical work on molecular reactivity and function will require in all but the simplest cases the study of a huge amount of molecular structures and their relation.
\item Automated procedures are mandatory for this task and recent developments have point into fruitful directions for such approaches.
\item Reaction-space exploration approaches must address a wide range of requirements ranging from stable exploration procedures for advanced atomistic modelling to uncertainty quantification for error assessment and cure to new visualization and immersion soft- and hardware to interact with complex networks of many thousand nodes.
\item The comparison of exploration algorithms and software is a mutli-dimensional task that requires to carefully assess the pros and cons regarding their theoretical background, efficient and stable implementation, and overall software engineering.
\end{enumerate}
\end{summary}

\begin{issues}[FUTURE ISSUES]
\begin{enumerate}
\item Future developments of algorithms for chemical (reaction) space exploration will require a very broad theoretical basis in order to cope with the many different scenarios that one may encounter in real-world molecular processes.
\item Such algorithmic developments will eventually demand a high level of sophistication of software engineering and integration in order to move on from advanced tools for specific problems to general tool boxes for explorations of molecular processes at the nanoscale.
\item Eventually, fully automated computational chemistry software will become a peer to a human operator with respect to molecular science in research and education if it can act autonomously on arguments and questions raised in natural language by the operator.
\end{enumerate}
\end{issues}

\section*{DISCLOSURE STATEMENT}
The authors are not aware of any affiliations, memberships, funding, or financial holdings that might be perceived as affecting the objectivity of this review. 

\section*{ACKNOWLEDGMENTS}
\label{sec:acknowledgments}
This work was financially supported by the Deutsche Forschungsgemeinschaft (DFG) (GZ: UN 417/1-1)
and by the Schweizerischer Nationalfonds (SNF) (Project 200021\_182400).

\begin{thebibliography}{135}
\expandafter\ifx\csname natexlab\endcsname\relax\def\natexlab#1{#1}\fi

\bibitem{Corey1995}
Corey EJ, Cheng XM. 1995.
The logic of chemical synthesis.
New York: Wiley, 1st ed.

\bibitem{Broadbelt1994}
Broadbelt LJ, Stark SM, Klein MT. 1994.
Computer generated pyrolysis modeling: on-the-fly generation of species,
  reactions, and rates.
\textit{Ind. Eng. Chem. Res.} 33:790--799

\bibitem{Broadbelt1996}
Broadbelt LJ, Stark S, Klein M. 1996.
Computer generated reaction modelling: decomposition and encoding algorithms
  for determining species uniqueness.
\textit{Comput. Chem. Eng.} 20:113--129

\bibitem{Broadbelt2005}
Broadbelt LJ, Pfaendtner J. 2005.
Lexicography of kinetic modeling of complex reaction networks.
\textit{AIChE J.} 51:2112--2121

\bibitem{Schwaller2018}
Schwaller P, Laino T, Gaudin T, Bolgar P, Bekas C, Lee AA. 2018{\natexlab{a}}.
Molecular transformer – a model for uncertainty-calibrated chemical reaction
  prediction.
\textit{ChemRxiv} :7297379.v2

\bibitem{Schwaller2018a}
Schwaller P, Gaudin T, Lányi D, Bekas C, Laino T. 2018{\natexlab{b}}.
“found in translation”: predicting outcomes of complex organic chemistry
  reactions using neural sequence-to-sequence models.
\textit{Chem. Sci.} 9:6091--6098

\bibitem{Segler2018}
Segler MHS, Preuss M, Waller MP. 2018.
Planning chemical syntheses with deep neural networks and symbolic {AI}.
\textit{Nature} 555:604--610

\bibitem{Pensak1977}
Pensak DA, Corey EJ. 1977.
{LHASA}--logic and heuristics applied to synthetic analysis, chap.~1.
 1--32

\bibitem{Ihlenfeldt1996}
Ihlenfeldt WD, Gasteiger J. 1996.
Computer-assisted planning of organic syntheses: The second generation of
  programs.
\textit{Angew. Chem. Int. Ed.} 34:2613--2633

\bibitem{Gothard2012}
Gothard CM, Soh S, Gothard NA, Kowalczyk B, Wei Y, et~al. 2012.
Rewiring chemistry: Algorithmic discovery and experimental validation of
  one-pot reactions in the network of organic chemistry.
\textit{Angew. Chem. Int. Ed.} 51:7922--7927

\bibitem{Kowalik2012}
Kowalik M, Gothard CM, Drews AM, Gothard NA, Weckiewicz A, et~al. 2012.
Parallel optimization of synthetic pathways within the network of organic
  chemistry.
\textit{Angew. Chem. Int. Ed.} 51:7928--7932

\bibitem{Segler2017a}
Segler MHS, Waller MP. 2017{\natexlab{a}}.
Neural-symbolic machine learning for retrosynthesis and reaction prediction.
\textit{Chem. Eur. J.} 23:5966--5971

\bibitem{Coley2017a}
Coley CW, Rogers L, Green WH, Jensen KF. 2017{\natexlab{a}}.
Computer-assisted retrosynthesis based on molecular similarity.
\textit{ACS Cent. Sci.} 3:1237--1245

\bibitem{Coley2017}
Coley CW, Barzilay R, Jaakkola TS, Green WH, Jensen KF. 2017{\natexlab{b}}.
Prediction of organic reaction outcomes using machine learning.
\textit{ACS Cent. Sci.} 3:434--443

\bibitem{Segler2017}
Segler MHS, Waller MP. 2017{\natexlab{b}}.
Modelling chemical reasoning to predict and invent reactions.
\textit{Chem. Eur. J.} 23:61180--6128

\bibitem{Coley2018}
Coley CW, Green WH, Jensen KF. 2018.
Machine learning in computer-aided synthesis planning.
\textit{Acc. Chem. Res.} 51:1281--1289

\bibitem{Simm2019}
Simm GN, Vaucher AC, Reiher M. 2019.
Exploration of reaction pathways and chemical transformation networks.
\textit{J. Phys. Chem. A} 123:385--399

\bibitem{Sameera2016}
Sameera WMC, Maeda S, Morokuma K. 2016.
Computational catalysis using the artificial force induced reaction method.
\textit{Acc. Chem. Res.} 49:763

\bibitem{Dewyer2018}
Dewyer AL, Arg{\"u}elles AJ, Zimmerman PM. 2018.
Methods for exploring reaction space in molecular systems.
\textit{WIREs Comput. Mol. Sci.} 8:e1354

\bibitem{Habershon2016}
Habershon S. 2016.
Automated prediction of catalytic mechanism and rate law using graph-based
  reaction path sampling.
\textit{J. Chem. Theory Comput.} 12:1786--1798

\bibitem{Kim2018}
Kim Y, Kim JW, Kim Z, Kim WY. 2018.
Efficient prediction of reaction paths through molecular graph and reaction
  network analysis.
\textit{Chem. Sci.} 9:825--835

\bibitem{Ismail2019}
Ismail I, Stuttaford-Fowler HBVA, Ochan~Ashok C, Robertson C, Habershon S.
  2019.
Automatic proposal of multistep reaction mechanisms using a graph-driven
  search.
\textit{J. Phys. Chem. A} 123:3407--3417

\bibitem{Bergeler2015}
Bergeler M, Simm GN, Proppe J, Reiher M. 2015.
Heuristics-guided exploration of reaction mechanisms.
\textit{J. Chem. Theory Comput.} 11:5712--5722

\bibitem{Simm2017}
Simm GN, Reiher M. 2017.
Context-driven exploration of complex chemical reaction networks.
\textit{J. Chem. Theory Comput.} 13:6108--6119

\bibitem{Grimmel2019}
Grimmel S, Reiher M. 2019.
The electrostatic potential as a descriptor for the protonation propensity in
  automated exploration of reaction mechanisms.
\textit{Faraday Discuss.} Accepted Manuscript, DOI: 10.1039/C9FD00061E

\bibitem{Rappoport2014}
Rappoport D, Galvin CJ, Zubarev DY, Aspuru-Guzik A. 2014.
Complex chemical reaction networks from heuristics-aided quantum chemistry.
\textit{J. Chem. Theory Comput.} 10:897--907

\bibitem{Rappoport2019}
Rappoport D, Aspuru-Guzik A. 2019.
Predicting feasible organic reaction pathways using heuristically aided quantum
  chemistry.
\textit{J. Chem. Theory Comput.} Accepted Manuscript, DOI:
  10.1021/acs.jctc.9b00126

\bibitem{Kim2014}
Kim Y, Choi S, Kim WY. 2014.
Efficient basin-hopping sampling of reaction intermediates through molecular
  fragmentation and graph theory.
\textit{J. Chem. Theory Comput.} 10:2419--2426

\bibitem{Habershon2015}
Habershon S. 2015.
Sampling reactive pathways with random walks in chemical space: Applications to
  molecular dissociation and catalysis.
\textit{J. Chem. Phys.} 143:094106

\bibitem{Ohno2004}
Ohno K, Maeda S. 2004.
A scaled hypersphere search method for the topography of reaction pathways on
  the potential energy surface.
\textit{Chem. Phys. Lett.} 384:277--282

\bibitem{Maeda2010}
Maeda S, Morokuma K. 2010.
Communications: A systematic method for locating transition structures of
  a+b→x type reactions.
\textit{J. Chem. Phys.} 132:241102

\bibitem{Maeda2013}
Maeda S, Ohno K, Morokuma K. 2013.
Systematic exploration of the mechanism of chemical reactions: the global
  reaction route mapping (grrm) strategy using the addf and afir methods.
\textit{Phys. Chem. Chem. Phys.} 15:3683--3701

\bibitem{Maeda2018}
Maeda S, Harabuchi Y, Takagi M, Saita K, Suzuki K, et~al. 2018.
Implementation and performance of the artificial force induced reaction method
  in the {GRRM17} program.
\textit{J. Comput. Chem.} 39:233--251

\bibitem{Zimmerman2015}
Zimmerman PM. 2015.
Single-ended transition state finding with the growing string method.
\textit{J. Comput. Chem.} 36:601--611

\bibitem{Dewyer2017}
Dewyer AL, Zimmerman PM. 2017.
Finding reaction mechanisms{,} intuitive or otherwise.
\textit{Org. Biomol. Chem.} 15:501--504

\bibitem{Wang2014}
Wang LP, Titov A, McGibbon R, Liu F, Pande VS, Mart{\'\i}nez TJ. 2014.
Discovering chemistry with an ab initio nanoreactor.
\textit{Nat. Chem.} 6:1044

\bibitem{Yang2017}
Yang M, Zou J, Wang G, Li S. 2017.
Automatic reaction pathway search via combined molecular dynamics and
  coordinate driving method.
\textit{J. Phys. Chem. A} 121:1351--1361

\bibitem{Huber1994}
Huber T, Torda AE, van Gunsteren WF. 1994.
Local elevation: A method for improving the searching properties of molecular
  dynamics simulation.
\textit{J. Comput.-Aided Mol. Des.} 8:695--708

\bibitem{Laio2002}
Laio A, Parrinello M. 2002.
Escaping free-energy minima.
\textit{Proc. Natl. Acad. Sci. U.S.A.} 99:12562--12566

\bibitem{Grimme2019}
Grimme S. 2019.
Exploration of chemical compound, conformer, and reaction space with
  meta-dynamics simulations based on tight-binding quantum chemical
  calculations.
\textit{J. Chem. Theory Comput.} 15:2847--2862

\bibitem{Rizzi2019}
Rizzi V, Mendels D, Sicilia E, Parrinello M. 2019.
Blind search for complex chemical pathways using harmonic linear discriminant
  analysis.
\textit{arXiv} :1904.06276v1

\bibitem{Martinez-Nunez2015}
Mart{\'i}nez-N{\'u}{\~n}ez E. 2015.
An automated method to find transition states using chemical dynamics
  simulations.
\textit{J. Comput. Chem.} 36:222--234

\bibitem{Varela2017}
Varela JA, V{\'a}zquez SA, Mart{\'i}nez-N{\'u}{\~n}ez E. 2017.
An automated method to find reaction mechanisms and solve the kinetics in
  organometallic catalysis.
\textit{Chem. Sci.} 8:3843--3851

\bibitem{Yang2018}
Yang M, Yang L, Wang G, Zhou Y, Xie D, Li S. 2018.
Combined molecular dynamics and coordinate driving method for automatic
  reaction pathway search of reactions in solution.
\textit{J. Chem. Theory Comput.} 14:5787--5796

\bibitem{Debnath2019}
Debnath J, Invernizzi M, Parrinello M. 2019.
Enhanced sampling of transition states.
\textit{J. Chem. Theory Comput.} 15:2454--2459

\bibitem{Grambow2018}
Grambow CA, Jamal A, Li YP, Green WH, Z{\'a}dor J, Suleimanov YV. 2018.
Unimolecular reaction pathways of a γ-ketohydroperoxide from combined
  application of automated reaction discovery methods.
\textit{J. Am. Chem. Soc.} 140:1035--1048

\bibitem{Maeda2019}
Maeda S, Harabuchi Y. 2019.
On benchmarking of automated methods for performing exhaustive reaction path
  search.
\textit{J. Chem. Theory Comput.} 15:2111--2115

\bibitem{Green1992}
Green WH, Moore CB, Polik WF. 1992.
Transition states and rate constants for unimolecular reactions.
\textit{Annu. Rev. Phys. Chem.} 43:591--626

\bibitem{Susnow1997}
Susnow RG, Dean AM, Green WH, Peczak P, Broadbelt LJ. 1997.
Rate-based construction of kinetic models for complex systems.
\textit{J. Phys. Chem. A} 101:3731--3740

\bibitem{Sumathi2002}
Sumathi R, Green~Jr. WH. 2002.
A priori rate constants for kinetic modeling.
\textit{Theor. Chem. Acc.} 108:187--213

\bibitem{Prinz2011}
Prinz JH, Wu H, Sarich M, Keller B, Senne M, et~al. 2011.
Markov models of molecular kinetics: Generation and validation.
\textit{J. Chem. Phys.} 134:174105

\bibitem{Sabbe2012}
Sabbe MK, Reyniers MF, Reuter K. 2012.
First-principles kinetic modeling in heterogeneous catalysis: an industrial
  perspective on best-practice{,} gaps and needs.
\textit{Catal. Sci. Technol.} 2:2010--2024

\bibitem{Bowman2013}
Bowman GR, Pande VS, No{\'e} F. 2013.
An introduction to markov state models and their application to long timescale
  molecular simulation, vol. 797.
Springer Science \& Business Media

\bibitem{VandeVijver2015}
Van~de Vijver R, Vandewiele NM, Bhoorasingh PL, Slakman BL, Seyedzadeh~Khanshan
  F, et~al. 2015.
Automatic mechanism and kinetic model generation for gas- and solution-phase
  processes: A perspective on best practices, recent advances, and future
  challenges.
\textit{Int. J. Chem. Kinet.} 47:199--231

\bibitem{Gao2016}
Gao CW, Allen JW, Green WH, West RH. 2016.
Reaction mechanism generator: Automatic construction of chemical kinetic
  mechanisms.
\textit{Comput. Phys. Commun.} 203:212 -- 225

\bibitem{Wu2016}
Wu H, Paul F, Wehmeyer C, No{\'e} F. 2016.
Multiensemble markov models of molecular thermodynamics and kinetics.
\textit{Proc. Natl. Acad. Sci. U.S.A.} 113:E3221--E3230

\bibitem{Proppe2016}
Proppe J, Husch T, Simm GN, Reiher M. 2016.
Uncertainty quantification for quantum chemical models of complex reaction
  networks.
\textit{Faraday Discuss.} 195:497--520

\bibitem{Doepking2017}
D{\"o}pking S, Matera S. 2017.
Error propagation in first-principles kinetic monte carlo simulation.
\textit{Chem. Phys. Lett.} 674

\bibitem{Han2018}
Han K, Green WH. 2018.
A fragment-based mechanistic kinetic modeling framework for complex systems.
\textit{Ind. Eng. Chem. Res.} 57:14022--14030

\bibitem{Vereecken2018}
Vereecken L, Aumont B, Barnes I, Bozzelli J, Goldman M, et~al. 2018.
Perspective on mechanism development and structure-activity relationships for
  gas-phase atmospheric chemistry.
\textit{Int. J. Chem. Kinet.} 50:435--469

\bibitem{Scherer2019}
Scherer MK, Husic BE, Hoffmann M, Paul F, Wu H, Noé F. 2019.
Variational selection of features for molecular kinetics.
\textit{J. Chem. Phys.} 150:194108

\bibitem{Andersen2019}
Andersen M, Panosetti C, Reuter K. 2019.
A practical guide to surface kinetic monte carlo simulations.
\textit{Front. Chem.} 7:202

\bibitem{Proppe2019}
Proppe J, Reiher M. 2019.
Mechanism deduction from noisy chemical reaction networks.
\textit{J. Chem. Theory Comput.} 15:357--370

\bibitem{Cavallotti2019}
Cavallotti C, Pelucchi M, Georgievskii Y, Klippenstein SJ. 2019.
Estoktp: Electronic structure to temperature- and pressure-dependent rate
  constants—a code for automatically predicting the thermal kinetics of
  reactions.
\textit{J. Chem. Theory Comput.} 15:1122--1145

\bibitem{Eyring1935}
Eyring H. 1935.
The activated complex in chemical reactions.
\textit{J. Chem. Phys.} 3:107--115

\bibitem{Kramers1940}
Kramers H. 1940.
Brownian motion in a field of force and the diffusion model of chemical
  reactions.
\textit{Physica} 7:284 -- 304

\bibitem{Haenggi1990}
H\"anggi P, Talkner P, Borkovec M. 1990.
Reaction-rate theory: fifty years after kramers.
\textit{Rev. Mod. Phys.} 62:251--341

\bibitem{Weymuth2014}
Weymuth T, Reiher M. 2014.
Inverse quantum chemistry: Concepts and strategies for rational compound
  design.
\textit{Int. J. Quantum Chem.} 114:823--837

\bibitem{Sanchez-Lengeling2018}
Sanchez-Lengeling B, Aspuru-Guzik A. 2018.
Inverse molecular design using machine learning: Generative models for matter
  engineering.
\textit{Science} 361:360--365

\bibitem{Freeze2019}
Freeze JG, Kelly HR, Batista VS. 2019.
Search for catalysts by inverse design: Artificial intelligence, mountain
  climbers, and alchemists.
\textit{Chem. Rev.} 119:6595--6612

\bibitem{Collins2014}
Collins KD, R{\"u}hling A, Glorius F. 2014.
Application of a robustness screen for the evaluation of synthetic organic
  methodology.
\textit{Nat. Prot.} 9:1348

\bibitem{Mayer1983}
Mayer I. 1983.
Charge, bond order and valence in the ab initio scf theory.
\textit{Chem. Phys. Lett.} 97:270 -- 274

\bibitem{Karpen1993}
Karpen ME, Tobias DJ, Brooks CL. 1993.
Statistical clustering techniques for the analysis of long molecular dynamics
  trajectories: analysis of 2.2-ns trajectories of ypgdv.
\textit{Biochemistry} 32:412--420

\bibitem{Shenkin1994}
Shenkin PS, McDonald DQ. 1994.
Cluster analysis of molecular conformations.
\textit{J. Comput. Chem.} 15:899--916

\bibitem{Jain1999}
Jain AK, Murty MN, Flynn PJ. 1999.
Data clustering: A review.
\textit{ACM Comput. Surv.} 31:264--323

\bibitem{Shao2007}
Shao J, Tanner SW, Thompson N, Cheatham TE. 2007.
Clustering molecular dynamics trajectories: 1. characterizing the performance
  of different clustering algorithms.
\textit{J. Chem. Theory Comput.} 3:2312--2334

\bibitem{Keller2010}
Keller B, Daura X, van Gunsteren WF. 2010.
Comparing geometric and kinetic cluster algorithms for molecular simulation
  data.
\textit{J. Chem. Phys.} 132:074110

\bibitem{Singhal2004}
Singhal N, Snow CD, Pande VS. 2004.
Using path sampling to build better markovian state models: Predicting the
  folding rate and mechanism of a tryptophan zipper beta hairpin.
\textit{J. Chem. Phys.} 121:415--425

\bibitem{Bowman2009}
Bowman GR, Huang X, Pande VS. 2009.
Using generalized ensemble simulations and markov state models to identify
  conformational states.
\textit{Methods} 49:197 -- 201

\bibitem{Li2016}
Li Y, Dong Z. 2016.
Effect of clustering algorithm on establishing markov state model for molecular
  dynamics simulations.
\textit{J. Chem. Inf. Model.} 56:1205--1215

\bibitem{Landrum2019}
Landrum G. 2019.
{RDKit: Open‐source cheminformatics}.
{http://www.rdkit.org}

\bibitem{Vaucher2016a}
Vaucher AC, Reiher M. 2016.
Molecular propensity as a driver for explorative reactivity studies.
\textit{J. Chem. Inf. Model.} 56:1470--1478

\bibitem{qcarchive}
{The QCArchive developers}. 2019.
{QCArchive}.
{https://qcarchive.molssi.org/}

\bibitem{Simm2017a}
Simm GN, Proppe J, Reiher M. 2017.
Error assessment of computational models in chemistry.
\textit{CHIMIA} 71:202--208

\bibitem{Simm2018}
Simm GN, Reiher M. 2018.
Error-controlled exploration of chemical reaction networks with gaussian
  processes.
\textit{J. Chem. Theory Comput.} 14:5238--5248

\bibitem{Wang2019}
Wang LP, Song C. 2019.
{C}ar-{P}arrinello monitor for more robust {B}orn-{O}ppenheimer molecular
  dynamics.
\textit{ChemRxiv} :8217362.v1

\bibitem{Vaucher2017}
Vaucher AC, Reiher M. 2017.
Steering orbital optimization out of local minima and saddle points toward
  lower energy.
\textit{J. Chem. Theory Comput.} 13:1219--1228

\bibitem{Hawkins2017}
Hawkins PCD. 2017.
Conformation generation: The state of the art.
\textit{J. Chem. Inf. Model.} 57:1747

\bibitem{Klamt1995}
Klamt A. 1995.
Conductor-like screening model for real solvents: A new approach to the
  quantitative calculation of solvation phenomena.
\textit{J. Phys. Chem.} 99:2224--2235

\bibitem{Grambow2019}
Grambow CA, Li YP, Green WH. 2019.
Accurate thermochemistry with small datasets: A bond additivity correction and
  transfer learning approach.
\textit{J. Phys. Chem. A} Accepted Manuscript, DOI: 10.1021/acs.jpca.9b04195

\bibitem{Sidler2016}
Sidler D, Schwaninger A, Riniker S. 2016.
Replica exchange enveloping distribution sampling (re-eds): A robust method to
  estimate multiple free-energy differences from a single simulation.
\textit{J. Chem. Phys.} 145:154114

\bibitem{Sidler2017}
Sidler D, Cristòfol-Clough M, Riniker S. 2017.
Efficient round-trip time optimization for replica-exchange enveloping
  distribution sampling (re-eds).
\textit{J. Chem. Theory Comput.} 13:3020--3030

\bibitem{SeveroPereiraGomes2012}
Severo Pereira~Gomes A, Jacob CR. 2012.
Quantum-chemical embedding methods for treating local electronic excitations in
  complex chemical systems.
\textit{Annu. Rep. Prog. Chem., Sect. C: Phys. Chem.} 108:222--277

\bibitem{Jacob2014}
Jacob CR, Neugebauer J. 2014.
Subsystem density-functional theory.
\textit{WIREs Comput. Mol. Sci.} 4:325--362

\bibitem{Wesolowski2015}
Wesolowski TA, Shedge S, Zhou X. 2015.
Frozen-density embedding strategy for multilevel simulations of electronic
  structure.
\textit{Chem. Rev.} 115:5891--5928

\bibitem{Lee2019}
Lee SJR, Welborn M, Manby FR, Miller TF. 2019.
Projection-based wavefunction-in-dft embedding.
\textit{Acc. Chem. Res.} 52:1359--1368

\bibitem{Husch2018}
Husch T, Vaucher AC, Reiher M. 2018.
Semiempirical molecular orbital models based on the neglect of diatomic
  differential overlap approximation.
\textit{Int. J. Quantum Chem.} 118:e25799

\bibitem{Parr1994}
Parr RG, Yang W. 1994.
Density-functional theory of atoms and molecules.
New York: Oxford University Press

\bibitem{Weymuth2014a}
Weymuth T, Couzijn EPA, Chen P, Reiher M. 2014.
New benchmark set of transition-metal coordination reactions for the assessment
  of density functionals.
\textit{J. Chem. Theory Comput.} 10:3092--3103

\bibitem{Simm2016}
Simm GN, Reiher M. 2016.
Systematic error estimation for chemical reaction energies.
\textit{J. Chem. Theory Comput.} 12:2762--2773

\bibitem{Husch2018a}
Husch T, Freitag L, Reiher M. 2018.
Calculation of ligand dissociation energies in large transition-metal
  complexes.
\textit{J. Chem. Theory Comput.} 14:2456--2468

\bibitem{Pernot2015}
Pernot P, Civalleri B, Presti D, Savin A. 2015.
Prediction uncertainty of density functional approximations for properties of
  crystals with cubic symmetry.
\textit{J. Phys. Chem. A} 119:5288--5304

\bibitem{Pernot2017}
Pernot P. 2017.
The parameter uncertainty inflation fallacy.
\textit{J. Chem. Phys.} 147:104102

\bibitem{Proppe2017}
Proppe J, Reiher M. 2017.
Reliable estimation of prediction uncertainty for physicochemical property
  models.
\textit{J. Chem. Theory Comput.} 13:3297--3317

\bibitem{Ma2018}
Ma Q, Werner HJ. 2018.
Explicitly correlated local coupled-cluster methods using pair natural
  orbitals.
\textit{WIREs Comput. Mol. Sci.} 8:e1371

\bibitem{Stein2016a}
Stein CJ, Reiher M. 2016.
Automated selection of active orbital spaces.
\textit{J. Chem. Theory Comput.} 12:1760--1771

\bibitem{Stein2016}
Stein CJ, von Burg V, Reiher M. 2016.
The delicate balance of static and dynamic electron correlation.
\textit{J. Chem. Theory Comput.} 12:3764--3773

\bibitem{Stein2019}
Stein CJ, Reiher M. 2019.
autocas: A program for fully automated multiconfigurational calculations.
\textit{J. Comput. Chem.} Accepted Manuscript, DOI: 10.1002/jcc.25869

\bibitem{Manby2012}
Manby FR, Stella M, Goodpaster JD, Miller TF. 2012.
A simple, exact density-functional-theory embedding scheme.
\textit{J. Chem. Theory Comput.} 8:2564--2568

\bibitem{Tamukong2014}
Tamukong PK, Khait YG, Hoffmann MR. 2014.
Density differences in embedding theory with external orbital orthogonality.
\textit{J. Phys. Chem. A} 118:9182--9200

\bibitem{Hegely2016}
H{\'e}gely B, Nagy PR, Ferenczy GG, Kállay M. 2016.
Exact density functional and wave function embedding schemes based on orbital
  localization.
\textit{J. Chem. Phys.} 145:064107

\bibitem{Muehlbach2018}
M{\"u}hlbach AH, Reiher M. 2018.
Quantum system partitioning at the single-particle level.
\textit{J. Chem. Phys.} 149:184104

\bibitem{Proppe2019a}
Proppe J, Gugler S, Reiher M. 2019.
Gaussian process-based refinement of dispersion corrections.
\textit{J. Chem. Theory Comput.} Submitted Manuscript

\bibitem{Haag2014}
Haag MP, Vaucher AC, Bosson M, Redon S, Reiher M. 2014.
Interactive chemical reactivity exploration.
\textit{ChemPhysChem} 15:3301--3319

\bibitem{Haag2014a}
Haag MP, Reiher M. 2014.
Studying chemical reactivity in a virtual environment.
\textit{Faraday Discuss.} 169:89--118

\bibitem{OConnor2018}
O{\textquoteright}Connor M, Deeks HM, Dawn E, Metatla O, Roudaut A, et~al.
  2018.
Sampling molecular conformations and dynamics in a multiuser virtual reality
  framework.
\textit{Sci. Adv.} 4:eaat2731

\bibitem{Amabilino2019}
Amabilino S, Bratholm LA, Bennie SJ, Vaucher AC, Reiher M, Glowacki DR. 2019.
Training neural nets to learn reactive potential energy surfaces using
  interactive quantum chemistry in virtual reality.
\textit{J. Phys. Chem. A} 123:4486--4499

\bibitem{Lu2017}
Lu Y. 2017.
The “ok, molly” chemistry.
\textit{Acc. Chem. Res.} 50:647--651

\bibitem{Aspuru-Guzik2018}
Aspuru-Guzik A, Lindh R, Reiher M. 2018.
The matter simulation (r)evolution.
\textit{ACS Cent. Sci.} 4:144--152

\bibitem{Fast2017}
Fast E, Chen B, Mendelsohn J, Bassen J, Bernstein M. 2017.
Iris: A conversational agent for complex tasks.
\textit{arXiv} :1707.05015v1

\bibitem{mycroft}
{The Mycroft developers}. 2019.
Mycroft.
{https://mycroft.ai/}

\bibitem{jarvis}
{The Jarvis developers}. 2019.
Jarvis.
{https://openjarvis.com/}

\bibitem{Lejaeghere2016}
Lejaeghere K, Bihlmayer G, Bj{\"o}rkman T, Blaha P, Bl{\"u}gel S, et~al. 2016.
Reproducibility in density functional theory calculations of solids.
\textit{Science} 351:aad3000

\bibitem{Glowacki2012}
Glowacki DR, Liang CH, Morley C, Pilling MJ, Robertson SH. 2012.
Mesmer: An open-source master equation solver for multi-energy well reactions.
\textit{J. Phys. Chem. A} 116:9545--9560

\bibitem{Kee1980}
Kee R, Miller J, Jefferson T. 1980.
Chemkin: a general-purpose, problem-independent, transportable, fortran
  chemical kinetics code package.
Tech. rep., Sandia Labs

\bibitem{Miller1993}
Miller WH. 1993.
Beyond transition-state theory: a rigorous quantum theory of chemical reaction
  rates.
\textit{Acc. Chem. Res.} 26:174--181

\bibitem{Truhlar1996}
Truhlar DG, Garrett BC, Klippenstein SJ. 1996.
Current status of transition-state theory.
\textit{J. Phys. Chem.} 100:12771--12800

\bibitem{Pollak2005}
Pollak E, Talkner P. 2005.
Reaction rate theory: What it was, where is it today, and where is it going?
\textit{Chaos} 15:026116

\bibitem{Garrett2005}
Garrett BC, Truhlar DG. 2005.
Theory and applications of computational chemistry, chap.~5.
Amsterdam: Elsevier,  67--87

\bibitem{FD2016}
 2017.
Reaction rate theory.
Faraday Discussions. The Royal Society of Chemistry

\bibitem{Richardson2018}
Richardson JO. 2018.
Perspective: Ring-polymer instanton theory.
\textit{J. Chem. Phys.} 148:200901

\bibitem{Reilly2016}
Reilly AM, Cooper RI, Adjiman CS, Bhattacharya S, Boese AD, et~al. 2016.
{Report on the sixth blind test of organic crystal structure prediction
  methods}.
\textit{Act. Cryst. B} 72:439--459

\bibitem{Rizzi2018}
Rizzi A, Murkli S, McNeill JN, Yao W, Sullivan M, et~al. 2018.
Overview of the sampl6 host--guest binding affinity prediction challenge.
\textit{J. Comput.-Aided Mol. Des.} 32:937--963

\bibitem{Maeda2012}
Maeda S, Morokuma K. 2012.
Toward predicting full catalytic cycle using automatic reaction path search
  method: A case study on hco(co)3-catalyzed hydroformylation.
\textit{J. Chem. Theory Comput.} 8:380--385

\bibitem{scine}
{The {SCINE} developers}. 2019.
{SCINE}: Software for chemical interaction networks.
{https://scine.ethz.ch/}

\end{thebibliography}

\end{document}